\documentclass[preprint]{aastex}

\begin{document}

\title{The Radio Galaxy Populations of Nearby Northern Abell Clusters}
\author{Neal A. Miller\altaffilmark{1,2,3}} 
\affil{National Radio Astronomy Observatory\altaffilmark{4}, P.O. Box O, \\ Socorro, New Mexico  87801}
\email{nmiller@aoc.nrao.edu}

\and

\author{Frazer N. Owen}
\affil{National Radio Astronomy Observatory\altaffilmark{4}, P.O. Box O, \\ Socorro, New Mexico 87801}
\email{fowen@aoc.nrao.edu}

\altaffiltext{1}{Visiting Astronomer, Kitt Peak National Observatory, National Optical Astronomy Observatories, which is operated by the Association of Universities for Research in Astronomy, Inc. (AURA) under cooperative agreement with the National Science Foundation.}

\altaffiltext{2}{Based in part on observations obtained with the Apache Point Observatory 3.5-meter telescope, which is owned and operated by the Astrophysical Reseach Consortium.}

\altaffiltext{3}{and New Mexico State University, Department of Astronomy, Box
30001/Dept. 4500, Las Cruces, New Mexico 88003}

\altaffiltext{4}{The National Radio Astronomy Observatory is a facility of the National Science Foundation operated under cooperative agreement by Associated Universities, Inc.}

\begin{abstract}
We report on the use of the NRAO VLA Sky Survey (NVSS) to identify radio galaxies in eighteen nearby Abell clusters. The listings extend from the cores of the clusters out to radii of 3$h_{75}^{-1}$Mpc, which corresponds to 1.5 Abell radii and approximately four orders of magnitude in galaxy density. To create a truly useful catalog, we have collected optical spectra for nearly all of the galaxies lacking public velocity measurements. Consequently, we are able to discriminate between those radio galaxies seen in projection on the cluster and those which are in actuality cluster members. The resulting catalog consists of 329 cluster radio galaxies plus 138 galaxies deemed foreground/background objects, and new velocity measurements are reported for 273 of these radio galaxies. 

The motivation for the catalog is the study of galaxy evolution in the cluster environment. The radio luminosity function (RLF) is a powerful tool in the identification of active galaxies, as it is dominated by star-forming galaxies at intermediate luminosities and active galactic nuclei (AGN) at higher luminosities. The flux limit of the NVSS allows us to identify AGN and star-forming galaxies down to star formation rates (SFR) less than 1M$_{\sun}$ yr$^{-1}$. This sensitivity, coupled with the all-sky nature of the NVSS, allows us to produce a catalog of considerable depth and breadth. In addition to these data, we report detected infrared fluxes and upper limits obtained from IRAS data.  It is hoped that this database will prove useful in a number of potential studies of the effect of environment on galaxy evolution.

\end{abstract}

\section{Introduction}

One of the major areas of research in extragalactic astronomy today is that of the effect of environment on the evolution of galaxies. The uniqueness of the cluster environment has been a major driver in such studies. Within a reasonable field size, clusters provide enough galaxies to draw statistically-significant conclusions over a broad range in local galaxy density. In addition, the diffuse intracluster medium (ICM) of clusters provides another influence on the evolution of member galaxies, as seen in the HI deficiency of cluster spirals and their often disturbed HI morphologies \citep{hayn1984}. 

Naturally, evolutionary studies have also examined the differences between clusters at moderate redshift with those seen nearby. \citet{butc1978} noted that the fraction of blue galaxies in clusters increases as one moves to higher redshifts. Presumably, this increase in blue fraction is the result of more active episodes of star formation in galaxy clusters at earlier epochs. While blue colors provide only a rough diagnostic of star formation, more accepted star formation indicators have been applied to clusters at low redshift. \citet{kenn1984} investigated the H$\alpha$ properties of several nearby clusters \citep[Cancer, Coma, A1367, and the Virgo cluster H$\alpha$ data from][]{kenn1983}. They tentatively concluded that the star formation rates (SFR) derived from H$\alpha$ were slightly lower in clusters than field galaxies. The uncertainty in this conclusion was due to the strong dependence of H$\alpha$ emission on Hubble type and the difficulty in assigning such classifications to the cluster galaxies. Furthermore, they identified several galaxies whose interesting properties were rarely seen outside of the cluster environment. These were generally faint, irregular galaxies with very strong H$\alpha$ emission and radio emission. They also noted that spiral galaxies in the core of Virgo seemed to have lower star formation rates.

Radio emission also happens to be an excellent probe of star formation. It is believed that as massive stars die in supernovae, they eject cosmic ray electrons which then spiral in the host galaxy's magnetic fields and emit synchrotron radiation \citep[see review in ][]{cond1992}. The lifetimes of such relativistic electrons are of the order $10^8$ years, and since they result from massive stars with short lifetimes themselves the radio emission is an indicator of nearly present star formation.\footnote{Note that optical indicators of SFR, in particular H$\alpha$, have timescales around $10^7$ years. However, the remarkable correlation of radio emission and H$\alpha$ emission seen for star-forming galaxies suggests that the overall duration of episodes of star formation makes this distinction minor.} Furthermore, the radio power seems to be directly related to the SFR. The radio luminosity function (RLF) at 1.4GHz represents a smooth continuum running from normal quiescent galaxies, to galaxies undergoing increasing levels of star formation, to relatively radio-weak AGN (e.g., Seyferts), and on to the rare yet extremely radio-powerful AGN \citep[the classic Fanaroff-Riley I and II sources;][]{fana1974}. The overall shape of the RLF closely resembles that of the more familiar optical luminosity function, and in each case the Milky Way lies very close to the knee in the LF.

In addition to the influence on star formation, numerous studies have investigated the effect of the cluster environment on active galactic nuclei (AGN), notably the powerful radio galaxies. The radio emission from these galaxies often extends well beyond their optical boundaries, out to hundreds of kiloparsecs. One might expect the ICM in clusters to confine the radio plasma from these galaxies, thereby affecting the total power of cluster radio galaxies. However, the RLF has not been shown to differ in clusters vs. the field \citep{fant1984, ledl1996}, nor does the probability for a galaxy to emit radio depend on cluster galaxy density \citep{zhao1989, leda1995}.

Consequently, a comprehensive catalog of nearby cluster radio galaxies is a useful tool in the study of the effect of the cluster environment on galaxy evolution. A large, comprehensive database will allow for more statistically robust studies of the above questions, plus detailed investigations of galaxy evolution in a range of cluster environments. 

In order to detect the more normal star-forming galaxies as well as the more radio-powerful AGN, sensitive radio maps are required. In the past, such sensitivity could be achieved reasonably easily for nearby clusters but the beam size of the radio observations restricted such studies to only the cores of such clusters. Technological advances have solved these problems. Lower noise radio receivers have reduced the time requirements for all-sky surveys at good sensitivity, and computing power has followed suit and made reduction of such a large volume of data feasible. Furthermore, procedures for creating near-uniform sensitivity mosaics of the radio data which accurately account for the problems inherent to wide-field radio imaging have been developed. One result of these advances is the NRAO VLA Sky Survey \citep[NVSS:][]{cond1998}, which covers the entire sky north of $\delta=-40$ at a frequency of 1.4 GHz. Performed with the VLA in D-array, it has a resolution of about 45$\arcsec$ and is sensitive enough to detect sources down to $S~\approx$ 2.5 mJy.

Within this paper, we use the NVSS to identify the radio galaxies in a sample of nearby clusters. The all-sky nature of the NVSS allows us to catalog radio galaxies out to great distances from the more well-studied cores of clusters. Adopting $H_o~=~75$ km s$^{-1}$ Mpc$^{-1}$ and $q_o~=~0.1$, we identify NVSS radio galaxies out to 3 Mpc from the cores of nearby clusters. This corresponds to approximately four orders of magnitude in galaxy density, thereby probing regions from the densest centers of clusters to those consistent with the average density of the field. The excellent sensitivity of the NVSS ensures that the identified galaxies include normal star-forming galaxies as well as powerful AGN. For example, at $z~=~0.015$ (about the nearest cluster in our sample) the flux limit of the NVSS corresponds to about $1 \times 10^{21}$ W Hz$^{-1}$, which means we detect objects with SFR$\gtrsim0.5$M$_\odot$ yr$^{-1}$ \citep[see][for the equation relating radio luminosity to SFR]{cari2001}. Even at our redshift cutoff the NVSS detects galaxies with SFR$\gtrsim3$M$_\odot$ yr$^{-1}$. The beam size of the NVSS is comparable to that of a spiral galaxy disk for the redshifts investigated, insuring that even diffuse radio emission spread over the disks of normal star-forming galaxies is detected.

In addition to the radio fluxes, we report both new and old data that are useful in investigating the effect of the cluster environment. We have compiled far infrared (FIR) data for all the radio galaxies, utilizing the IRAS Faint Source and Point Source Catalogs (FSC and PSC, respectively). In the event such data are not available, we have used IPAC's {\it xscanpi} software to estimate fluxes from cross scans of the IRAS satellite. Perhaps most importantly, we report velocities for all but two of the identified radio galaxies, with many of these velocities arising from new observations. These data differentiate the true cluster radio objects from those merely seen in projection on the cluster. 

The paper is organized as follows. In section two, we present the cluster sample and summarize the procedure we have used to identify the candidate radio galaxies in each cluster. The compilation of public data, including velocities and FIR fluxes, is also discussed in this section. In section three, we discuss the spectral observations we have made for many of the candidate radio galaxies. This includes a summary of where, when, and how the data were collected, and our reduction and analysis procedures. Section four presents the results, in the form of complete tables for both the confirmed cluster radio galaxies and those radio galaxies deemed foreground or background objects. In section five we conclude with some simple conclusions and an overview of potential uses for the data.

\section{Sample Identification}

\subsection{Cluster Sample}

The galaxy clusters analyzed are taken from the catalog of \citet{abel1989} (ACO). This revision of Abell's original catalog \citep{abel1958} contains 4073 potential clusters of galaxies, as determined from inspection of the Palomar Observatory Sky Survey (POSS I) plates and the Southern Sky Survey. We have imposed a systemic recessional velocity cutoff of 10,000 km s$^{-1}$ ($z=0.033$), as beyond this redshift the sensitivity of the NVSS would no longer allow us to detect galaxies with massive SFRs below about 2 M$_\odot$ yr$^{-1}$. In addition, we have accepted only clusters with positive declinations although the NVSS coverage extends down to $\delta\sim-40$. While there are a large number of clusters meeting these criteria, many possess fairly faint magnitudes for their tenth-brightest member despite supposedly low systemic redshifts, making them somewhat uncertain. Furthermore, the areal coverage required to survey radial extents of 3 Mpc becomes prohibitively large at low redshift. As a result, we have settled on the 18 ACO clusters presented in Table \ref{tbl-1}.

\placetable{tbl-1}

The sample spans a range of cluster environments. Half of the members are poor clusters of richness class 0 \citep[see][]{abel1958}, while the other half are fairly rich systems. In fact, four of the clusters are of richness class 2, which means that they are richer than approximately 95$\%$ of all clusters. There are examples of cooling flow clusters (A426 and A2199), clusters with features suggesting recent large-scale interactions such as cluster-group mergers (A400, A1656, and A2634), and a number of clusters which together make up superclusters.

\subsection{Generating Optical Galaxy Lists}

The optical identifications are based on the Second Palomar Observatory Sky Survey \citep[POSS II:][]{reid1991}. The overwhelming success of the POSS I coupled with advances in photographic emulsion led to the initiation of this study, which offers both improved resolution and deeper limiting magnitudes than the original survey (1$\arcsec~pixel^{-1}$ vs. 1.7$\arcsec~pixel^{-1}$, and $R_c(lim) \sim 21$ vs. $R_c(lim) \sim 20$, respectively). We have used the IIIa-F images ($\lambda_{eff}\sim6500\mbox{\AA}$, bandwidth $\sim1000\mbox{\AA}$) in their digitized form\footnote{The Digitized Sky Surveys were produced at the Space Telescope Science Institute under U.S. Government grant NAG W-2166. The images of these surveys are based on photographic data obtained with the Oschin Schmidt Telescope on Palomar Mountain and the UK Schmidt Telescope. The plates were processed into the present compressed digital form with the permission of these institutions.} for our optical identifications, positions, and approximate R-band magnitudes.

For each cluster, we adopt a center position and system redshift (see Table \ref{tbl-1}). Then, a grid of overlapping square maps sufficient to cover the entire cluster out to 3 Mpc in radial distance from the assigned center are obtained. Each image then has a coordinate system applied using the plate solutions found in the headers of the FITS images, and the resulting coordinates are generally accurate to about 1$\arcsec$ \citep{deut1999}. SAD (``Search And Destroy''), a task within NRAO's Astronomical Image Processing System (AIPS), is then used to identify all objects significantly above the noise level on POSS II images. As stars may be represented by (often saturated) point source profiles and galaxies are fainter, more extended objects, we make a strong first cut at galaxy/star segregation using the source fluxes and widths reported by SAD. SAD also reports the centers of the identified objects, as determined by the best-fitting Gaussian to the real data.

We then calculate the equivalent R-band magnitudes for the identified galaxies. All our magnitudes are calculated for the Gunn-Oke aperture \citep{gunn1975}, which corresponds to a linear aperture radius of 13.1 kpc (for $H_o~=~75$ km s$^{-1}$ Mpc$^{-1}$) at the systemic redshift of the cluster. To establish the magnitude zero point, we have bootstrapped the POSS II images to the photometry reported in \citet{ledl1995}. This study reported photometric results for the powerful radio galaxies in ACO clusters, and as a result there is usually a galaxy or two in each of our clusters that has a published magnitude from which to work. In the event that there is no galaxy in common between the Ledlow \& Owen photometry and our clusters, we have adopted a magnitude zero point of 30.10, about the global average for the common galaxies in the entire sample. The adopted magnitude zero point for each cluster may also be found in Table \ref{tbl-1}. Once these parameters are established, the aperture photometry is preformed automatically using the PHOT task in IRAF. The local background for each galaxy is calculated using a large annulus centered on the galaxy. The subsequent magnitudes are generally accurate to within 0.5 magnitudes, with this error representing roughly equal contributions from uncertainty in the magnitude zero point and photometric measurement.

To remove probable background objects and reduce the galaxy list to those most likely to be associated with each cluster, we discard the fainter objects. Our cutoff magnitude is $m^*+2$, where we have assumed that $M_{R}^*=-22$ \citep{owen1989} and the attendant $m^*$ is calculated based on the systemic velocity of each cluster. In addition, we allow for Galactic extinction in the direction of the cluster (see Table \ref{tbl-1}). 

\subsection{Identification of Radio Galaxies}

For any given radio source, there is a definable probability that it will randomly be associated with an optical galaxy. This probability is a function of radial separation, such that perfect positional coincidence has a very high probability of representing true association whereas larger separations are more likely to be the result of common chance. The probability that a radio source is randomly associated with an optical counterpart within separation $r$ is

\begin{equation}
P(<r)~=~1~-~\exp(-\rho \pi r^2)
\end{equation}

\noindent
where $\rho$ is the density of background objects (i.e., the average density of optical galaxies about the radio objects).

In creating our list of radio galaxies, we have imposed a probability of $0.5\%$, or statistically, we expect that fewer than 1 out of every 200 identified radio galaxies is the result of chance superposition. To determine $\rho$ we take the NVSS catalog list of radio sources in the area surveyed for each cluster and calculate the average density of objects from our optical galaxy list about each radio source. Given these parameters we solve Equation 1 for the search radius to use in each cluster. As this is done for each cluster, the search radius from cluster to cluster is a variable but is generally in the range of 15$\arcsec$ to 35$\arcsec$. Then, if the nearest optical galaxy for a given radio source is within this radius we associate the pair as a radio galaxy.

While many studies simply use a fixed radio and optical separation to identify radio galaxies, we feel the probabilistic approach is better suited to this study. First, since the clusters are at different redshifts a fixed angular separation results in a more restrictive acceptance criterion for nearby galaxies. Optically, the nearer galaxies in the sample can easily span two arcminutes, whereas the farthest galaxies are well under an arcminute in extent. Relative errors in the galaxy center positions are therefore more likely to be larger for nearby galaxies, so a larger search radius is desirable. More importantly, our procedure consistently handles variations from field to field. For example, any errors in astrometric registration are accounted for in finding the search radius for that specific field.

After these automated routines are completed, the entire field for each cluster is inspected by eye. NVSS images are obtained, and overlays of NVSS contours on the POSS II images are created and inspected. This step is time consuming but serves a number of important functions. First, the galaxy/star identification algorithm is not perfect and a number of apparent close binary stars are labelled galaxies. These objects may be easily identifed and removed from the optical galaxy list during visual inspection. Second, any galaxy that was missed for some reason by the automated procedures may be added in by hand. This type of error rarely occurred except in the cases of very large galaxies. Once these corrections have been made, the updated optical list and the NVSS Catalog list are again correlated with one another to determine the semi-final list of radio galaxies for each cluster. In this step, the search radius is recalculated to reflect the updates in the list of optical galaxies.

The final purpose of the visual inspection is the identification of powerful, extended radio sources. Since the NVSS Catalog was generated by an automated procedure which fit peaks in radio emission, extended sources are frequently fit by multiple sources with smaller fluxes. There are two net effects of this on our radio galaxy identification procedure. First, the fluxes for some of our radio galaxies will be under-represented by their fluxes in the NVSS Catalog should the radio source be resolved into multiple components. For these, we have used the actual NVSS images to measure the net flux associated with the radio galaxy. Second, for some sources the fitted peak in the NVSS Catalog lies outside our adopted search radius (for example, see 3C465 in A2634). As these powerful radio galaxies are an important portion of our radio galaxy catalog, it is vital that they be identified. Table \ref{tbl-2} lists all of the identified candidates for such inclusions along with the separations between our optical positions and the nearest NVSS Catalog radio position, their net fluxes as measured off the NVSS images, and their common name where applicable. Overlays of the NVSS contours on the POSS II images for candidate extended radio sources are presented in Figure \ref{fig1}.

\placetable{tbl-2}

\placefigure{fig1}

As the resolution of the NVSS is about 45$\arcsec$, we have used Faint Images of the Radio Sky at Twenty Centimeters \citep[FIRST,][]{beck1995} to investigate the extended sources whenever such images were available. FIRST is also being undertaken with the VLA, using the same pointing grid and mosaic strategy as the NVSS but in the longer baseline B-array. As such, it provides a resolution of about 5$\arcsec$ with 5$\sigma$ detections near 1 mJy. The smaller beam size makes it less sensitive to diffuse, extended sources yet it is useful in demonstrating that some of our apparent extended NVSS radio galaxies are chance projections of more compact objects at higher redshift. Figure \ref{fig1} presents the FIRST images for the candidate extended sources, and Table \ref{tbl-2} indicates whether such sources were accepted in the final list of radio galaxies.

This procedure was applied to all the clusters, with one minor exception. The NVSS radio image of A426, the Perseus cluster, is dominated by the extremely powerful radio galaxy NGC1275, which is also known as Perseus A or 3C84. Since the core of this source is unresolved and at a radio power of nearly 25 Jy, there are significant artifacts in the central region of this cluster. Therefore, we have used the radio galaxy identifications of \citet{sijb1993} within 0.5$h_{75}^{-1}$Mpc of Perseus A. The radio galaxies chosen in this way were required to be consistent with the rest of this study, in that we accepted only the Sijbring radio galaxies with fluxes which should render them detected by the NVSS.

\section{Characterization of Radio Galaxies}

\subsection{Infrared Data}

A useful characterization of galaxies with radio emission is provided by their far infrared (FIR) emission. For normal star-forming galaxies as well as those actively involved in starbursts, the radio and FIR fluxes are strongly correlated \citep[see review in][]{cond1992}. While the two types of emission are different (non-thermal for the radio and thermal for the FIR), the root cause in each case is massive stars. In the radio, as these massive stars die in supernovae they accelerate electrons which spiral in the galaxy's magnetic fields and emit synchrotron radiation. During their brief lifetimes, the massive stars are the primary heaters of dust in HII regions. As a result, galaxies which are presently forming massive stars follow the FIR-radio correlation.

A useful measure of the FIR-radio correlation is the statistic $q$, defined by \citet{helo1985} as

\begin{equation}
q~\equiv~\log \left( \frac{FIR}{3.75\times 10^{12}W~m^{-2}} \right) - \log \left( \frac{S_{1.4GHz}}{W~m^{-2}~Hz^{-1}} \right)
\end{equation}

\noindent
where FIR is defined as

\begin{equation}\label{eqn:fir}
\left( \frac{FIR}{W~m^{-2}} \right) ~\equiv~ 1.26 \times 10^{-14} \left( \frac{2.58 S_{60\mu m}+S_{100\mu m}}{Jy} \right) .
\end{equation}

\noindent
The numerical terms in the FIR expression are used to convert the IRAS flux densities into a representation of total flux between about 40$\mu$m and 120$\mu$m. Thus, $q$ is a logarithmic measure of the FIR/radio flux density ratio. For a wide variety of star-forming galaxies (including E's and S0's with active star formation), the distribution of $q$ is very narrow. The median value is about 2.3, with a standard deviation of around 0.2. Consequently, $q$ may be used to characterize radio galaxies as either star-forming or AGN. In AGN, there is effectively an excess of radio emission relative to FIR emission leading to low values of $q$.

We have used the NASA/IPAC Extragalactic Database (NED)\footnote{The NASA/IPAC Extragalactic Database (NED) is operated by the Jet Propulsion Laboratory, California Institute of Technology, under contract with the National Aeronautics and Space Administration.} to collect IRAS data on all galaxies in the sample. NED contains the information from the IRAS Point Source and Faint Source Catalogs (PSC and FSC, respectively), so galaxies with such FIR fluxes will be so marked. However, lack of a NED IRAS entry does not necessarily imply that the galaxy in question was not detected by IRAS. Some galaxies with detected IR emission were left out of the catalogs on account of their proximity to other IR sources. 

For all galaxies lacking NED FIR entries, we have used IPAC's {\it xscanpi} software to estimate the 60$\mu$m and 100$\mu$m fluxes. {\it xscanpi} performs coaddition of all IRAS scans across a given position and returns information such as peak flux, integrated flux, and rms noise. It is thereby useful in determining fluxes for confused or faint sources, and for determining upper limits to FIR fluxes. The data are presented for four different averaging schemes, corresponding to weighted mean, straight mean, median, and noise-weighted mean. We have adopted the median results as they are purported to be ``the most consistently `good' estimator... chiefly because of the non-Gaussian nature of noise in the IRAS data."\footnote{\url{http://www.ipac.caltech.edu/ipac/iras/scanpi\_interp.html}} The flux estimates we adopt correspond to a spatial bandpass equivalent to that of the IRAS resolution at each 60$\mu$m and 100$\mu$m.

In addition to the radio galaxies lacking NED FIR entries, around 10$\%$ of the radio galaxies are IRAS Point Source Catalog (PSC) objects with 60$\mu$m detections and 100$\mu$m upper limits (90$\%$ confidence). We have also used {\it xscanpi} to re-evaluate the 60$\mu$m and 100$\mu$m flux densities of these galaxies, as the cross scan addition reduces the noise and can provide firmer 100$\mu$m flux densities. Figure \ref{fig2} depicts the $q$ values determined from the PSC catalog compared with their corresponding cross scan addition results. As expected, the $q$ values show a near one-to-one correspondance but with the PSC-derived upper limit values producing slightly higher $q$. In general, the addscan fluxes for sources detected at both 60$\mu$m and 100$\mu$m are consistent with the PSC fluxes to within a few percent \citep{saun2000}. The addscan fluxes are on average slightly higher because the scan addition counts interacting systems as single sources whereas they may be separated into distinct components in the PSC.

\placefigure{fig2}

Using these values for the 60$\mu$m and 100$\mu$m fluxes in conjunction with the NVSS radio data, we calculated $q$ values for all of the identified radio galaxies. It should be noted that these $q$ values frequently represent upper limits on account of uncertainty in the IR fluxes. Nevertheless, they may be used in conjunction with the absolute radio powers of the sources to obtain a good idea of the nature of the radio source, star-forming galaxy or AGN. Figure 3 demonstrates this in the form of a histogram of $q$ values for the confirmed cluster members (see next section). It can easily be seen that the distribution of $q$ values peaks at 2.3, and consequently that the radio emission for the majority of the identified galaxies originates from star formation. This statement is valid in the statistical sense, in that the FIR and radio emission of AGN are also correlated but exhibit a greater scatter in their $q$ values. Thus, while most AGN have $q\lesssim2$ their distribution includes some galaxies with $q$ values representative of normal star-forming galaxies. Conversely, some star-forming galaxies in clusters are radio overluminous (i.e., lower $q$). These topics are addressed in \citet{mill2001}.

\placefigure{fig3}

\subsection{Velocity Data}

In order to confirm or refute the cluster membership of the identified radio galaxies, velocity measurements are necessary. We have relied upon the published velocity dispersions of \citet{stru1999} to assess cluster membership. These velocity dispersions were calculated using all publicly-available redshift measurements as of December 1998, and are listed along with the other adopted cluster parameters in Table \ref{tbl-1}. Galaxies whose velocities differ from the adopted cluster systemic velocity by less than three times the adopted velocity dispersion were considered to be cluster members. Those failing this criterion were designated non-cluster members, and are assumed to be either foreground or background objects. 

In searching NED for IR data on the galaxies, we have also noted the redshifts whenever such data were available. However, due to the large angular extent of our study velocity data were not available for many radio galaxies. In addition, many of the radio galaxies were deemed interesting for specific science questions that required optical spectra. Consequently, we have obtained long-slit optical spectra for many of the galaxies within the sample.

\subsubsection{Optical Spectroscopy Observations}

The spectra were obtained during the course of several observing runs at Apache Point Observatory (APO) and Kitt Peak National Observatory (KPNO). The APO observations were made using the ARC 3.5-meter telescope with the Double Imaging Spectrograph (DIS), while the KPNO observations were made using the 2.1-meter telescope with the GoldCamera Spectrograph (GCAM). In addition to the objective of determining accurate radial velocities, we also desired high signal-to-noise spectra for line ratio diagnostics and other analyses. Consequently, we employed moderate resolution gratings which provided wavelength coverage of the entire optical window and at a resolution sufficient to resolve line complexes such as H$\alpha$ and [NII]. Dates and relevant parameters of the observations are provided in Table \ref{tbl-3}.

\placetable{tbl-3}

Thanks to the large range in right ascension of the sample, we were always able to observe the target radio galaxies at low airmass. As a result of this and to increase our efficiency in collecting the spectra, we performed no slit rotations. In each observation, we simply aligned the slit with the nucleus of the target galaxy. Once the target was so acquired, the duration of the exposure was usually fifteen minutes. In some instances, we reduced the exposure time for particularly bright galaxies or lengthened the exposure times when there was suspected cirrus.

\subsubsection{Spectroscopic Data Reduction and Velocity Measurement}

The data were reduced in the usual manner using IRAF. The two-dimensional spectra were overscan subtracted, bias subtracted, and flat fielded using stacked frames of calibration quartz lamps, with twilight frames being used to correct for the illumination pattern of the slit. The wavelength scales were derived using observations of HeNeAr lamps. To account for telescope flexure and temperature changes throughout the nights, at least one set of calibration lamps was taken at each cluster. The long slits provided plenty of area to perform accurate sky subtraction.

Velocity measurement followed one of two paths. Since many of the galaxies are star forming, they possessed a number of emission lines. Whenever present, we measured the velocities using these lines. Line locations were determined using Gaussian fits, with each detected emission line receiving equal weight in the determination of the galaxy's recessional velocity and associated uncertainty. The detected lines are listed along with the resultant redshifts in Table \ref{tbl-4}. Repeated observations of velocity standard stars on each evening of the run showed agreement within about 20 km s$^{-1}$. 

For many of the galaxies, emission lines were not present or were too noisy to be definitive in velocity measurement. Additionally, we required more than one emission line (or complex, such as H$\alpha$--[NII]) to report an emission line velocity. In these cases, we have used the IRAF task FXCOR. This task utilizes a fourier cross correlation procedure \citep{tonr1979} to match a template absorption line spectrum with known heliocentric velocity to the target spectrum. We used the radial velocity standard stars HD132737 and HD090861, obtained during the first and second KPNO observing runs  respectively, as our template spectra. We required a fit which produced $R$ greater than five \citep[see][]{tonr1979}, though in nearly all cases the correlation was much better than this and typically $R$ was around twenty. The resulting heliocentric redshifts and associated errors are presented in Table \ref{tbl-4}. Measurements determined via this cross correlation technique are those for which either no emission lines or emission lines in parentheses are listed in Table \ref{tbl-4}.

\placetable{tbl-4}

\section{Results}

The total number of radio galaxies identified by the procedures outlined above was 467. Of these, 329 were deemed cluster members by their measured velocities, including three galaxies for which no accurate velocity measurement was obtained. These radio galaxies are listed by cluster in ascending order of right ascension in Table \ref{tbl-5}. In addition to the optical positions of the radio galaxies, a number of valuable parameters are provided. These include the redshift, R-band magnitude, radio and infrared flux densities (from the NVSS and IRAS, respectively), and $q$ value. Table \ref{tbl-6} presents the same information for the non-cluster radio galaxies.

\placetable{tbl-5}

\placetable{tbl-6}

A few things should be noted about the values presented in Tables \ref{tbl-5} \& \ref{tbl-6}. The coordinates for the radio galaxies (columns 2 and 3) correspond to the optical centers of the host galaxies, as measured from the POSS II images. These measured positions correspond to the peak of the best-fit Guassian to the galaxies, and as noted the astrometric solutions for the images are generally only good to within about 1.5$\arcsec$\citep{deut1999}. For those galaxies for which we have not obtained an optical spectrum, the listed redshifts were obtained from NED. The reader is directed to this resource for specific references for these values as well as their attendant errors. As noted in Section 2, the apparent magnitudes presented in column 5 should be accurate to within 0.5 magnitudes. As they are calculated using the Gunn-Oke aperture, they fall short of the full galaxy magnitude for the larger galaxies. Some of the FIR fluxes reported in columns 7 and 8 should be taken with caution. Our reliance on the median averaging in {\it xscanpi} combined with the flux estimator assuming a fixed signal range means we sometimes get detected fluxes for sources which are in actuality only nearby a true source. These sources can usually be identified as those within a few arcminutes of one another, and their $q$ values should be viewed in this light. Lastly, there are no IRAS scans in the direction of A1267 and hence there are no reported FIR flux densities or $q$ values for this cluster.

Simple conclusions may be drawn from Table \ref{tbl-5}. As would be expected, the four clusters possessing the most radio galaxies are the richness class 2 clusters. A426 (Perseus) has 43 confirmed radio galaxies, making it the richest system in the sample. However, the proximity of A426 coupled with the flux-limited radio data contribute strongly to this result. Applying a uniform radio luminosity cut-off makes A1656 (Coma) as rich as A426 in radio galaxies. Several of the richness class 0 clusters have very few radio galaxies and likely represent chance superpositions of unrelated systems. For example, 10 out of 17 identified radio galaxies in the field of A1267 turned out to be background objects. The distributions of the radio galaxies in individual clusters also appear to reflect larger scale structures, often aligning with nearby clusters or including distinct sub-clumps within the clusters.

A number of interesting things about the non-cluster radio galaxies can be seen from Table \ref{tbl-6}. These galaxies are frequently clumped in velocity space, indicating the presence of background clusters which may or may not be associated with Abell clusters. In some instances, it seems likely that cluster identified by Abell strictly on the basis of magnitudes represents the superposition of potentially unrelated systems. For example, A2162 contains only four radio galaxies consistent with the cluster at $z\sim0.03$, whereas an additional six radio galaxies are at $z\sim0.05$ (sometimes referred to as A2162N). In other cases, our identification procedure has resulted in the inclusion of background Abell clusters - such as A1213, concentrated in two clumps located in the northeast and southeast of A1185. While such systems may trace larger-scale structures, it appears that the majority of the galaxies labelled non-cluster are background radio galaxies. Only 18 of the 138 non-cluster radio galaxies lie between three and five times the velocity dispersion of their projected cluster, with most of the remaining 120 at much greater velocities.

As might be expected, many of these background radio sources are also detected in surveys at other frequencies such as the 87GB Catalog \citep{greg1991}. As the galaxies apparently associated with these radio sources formally met our selection criteria, we obtained optical spectra for them and confirmed that they are, in fact, powerful radio galaxies at higher redshift. Included among these are several galaxies which lay just outside our 3$h_{75}^{-1}$Mpc search radius and are consequently left out of Table \ref{tbl-6}, but may be found in the spectroscopy results of Table \ref{tbl-3}. For example, the source 113049+252436 was found to be at a redshift of 0.1444 and consequently have a radio power of $10^{24.5}$ W Hz$^{-1}$.

We also note a few cases where powerful known radio galaxies lie just outside the 3$h_{75}^{-1}$Mpc limit of our sample but at redshifts consistent with cluster membership. 3C76.1 is at a projected separation of $\sim 3.3h_{75}^{-1}$Mpc from the center of A397 and at the same redshift as the cluster. B2 1108+27 also seems to be associated with A1185, lying $\sim3.5h_{75}^{-1}$Mpc from that cluster's core and at essentially the same redshift. Lastly, we find that UGC02755 appears to be associated with a powerful radio source. This galaxy is barely excluded from our sample, as it is 3.01$h_{75}^{-1}$Mpc from the center of A426. The presence of powerful radio galaxies such as these at significant distances from the centers of these Abell clusters is strongly suggestive of the importance of large-scale structure in the formation and evolution of radio galaxies.

In addition to the summary information presented in Table \ref{tbl-5}, we have computed absolute quantities for the cluster radio galaxies. In these calculations we have assumed that all of the galaxies within a cluster lie at the same distance, which was determined from the cluster systemic velocity adopted in Table \ref{tbl-1}. Hence, we assume that the galaxies are bound cluster members and that the redshift range within each cluster is simply the result of peculiar velocities. Table \ref{tbl-7} presents the absolute R-band magnitude, 1.4 GHz radio power, and FIR luminosity (derived based on Equation \ref{eqn:fir}) for each cluster radio galaxy.

The utility of the absolute quantities is perhaps greatest for investigation of the SFRs of the cluster galaxies. Typically, the 1.4 GHz radio luminosity is converted to a SFR using the relationship found in \citet{cond1992}. This relationship is derived empirically using the radio luminosity and supernova rate of the Milky Way, and uncertainty in such quantities translates to uncertainty in derived SFRs. In comparison to other wavelength estimators of SFR, the radio determination likely produces higher SFRs \citep[e.g.,][find the radio SFR to be several times greater than extinction-corrected optical SFRs]{haar2000}. An alternate derivation of the relationship between radio luminosity and SFR which yields values more in line with optical results is presented in \citet{cari2001}. In the current paper, we have adopted this relationship. Using the flux limit of the NVSS for the range in redshift covered in this catalog, galaxies with SFR$\gtrsim0.5-2.7$M$_\odot$ yr$^{-1}$ are detected. 

\placetable{tbl-7}

\section{Conclusions}

We have presented a comprehensive catalog of the radio galaxies in nearby Northern Abell clusters. This catalog is complete down to relatively low radio powers, and our collection and compilation of velocity data ensures that the true cluster members are separated from those radio galaxies viewed in projection. In conjunction with these data, we report FIR fluxes and corresponding values of $q$, a parametrization of the FIR-radio correlation. Using the cluster velocities, we have compiled a list of the absolute magnitudes, radio powers, and FIR luminosities of the confirmed cluster radio galaxies.

\acknowledgments

This paper is the result of many nights of spectroscopic observations. We would like to thank the instrument specialists and telescope operators who were tremendously helpful in making sure these evenings went smoothly - Jim De Veny, Hal Halbedel, Heidi Schweiker, Doug Williams, Eugene McDougall, Karen Loomis, Camron Hastings, and Russet McMillan. N.A.M. would like to thank the NRAO pre-doctoral program for support of this research.

\clearpage

\figcaption{Extended radio galaxies added to the sample (see Table 2). The NVSS positions of these galaxies were formally outside the search radius for their parent clusters, yet they were included as possible cluster radio galaxies on the basis of their morphologies. The optical images are all taken from the POSS II, and source of the overlaid contors is given for each plot. The base contour level for the NVSS is 450$\mu$Jy, with contours at 2, 5, 10, 20, 40, 80, 160, and 360. For FIRST, the base contour level is 150$\mu$Jy with contours at 5, 10, 20, 40, 80, 160, and 320. (a) Page one, top left - 022259+430042, NVSS contours. This source appears to exist in its own subpeak within the greater extended emission of 3C66B; (b) Page one, top right - 022308+412211, 022311+412204, NVSS contours. Each of these sources were accepted as radio galaxies; (c) Page one, bottom left - 025654+164844, NVSS contours. (d-e) Page two, top left and top right - 125635+281628, NVSS contours and FIRST contours, respectively. Rejected as the FIRST data show this is likely a background source; (f-g) Page two, bottom left and bottom right - 125818+290739, NVSS contours and FIRST contours, respectively. (h-i) Page three, top left and top right - 161829+295859, NVSS contours and FIRST contours, respectively. (j-k) Page three, bottom left and bottom right - 162505+394529, NVSS contours and FIRST contours, respectively. (l-m) Page four, top left and top right - 162842+400725, NVSS contours and FIRST contours, respectively. On the basis of the NVSS image, both of these galaxies were accepted. The emission is resolved out by FIRST. (n) Page four, bottom left - 233839+270040, NVSS contours. The galaxy can be seen to reside in its own subpeak within 3C465's extended emission. (o) Page four, bottom right - 234022+271104, NVSS contours. \label{fig1}}

\figcaption{$q$ values derived from {\it xscanpi} vs. those derived from the IRAS Point Source Catalog (PSC). All galaxies on this plot were ones where the PSC reported a $90\%$ confidence upper limit on the 100$\mu$m flux density. Filled circles represent cases where the cross scans yielded 100$\mu$m detections of such galaxies at greater than 3$\sigma$, whereas open circles remained upper limits. The solid line demonstrates that the two evaluations of $q$ are consistent with one another. \label{fig2}}

\figcaption{Histogram of $q$ values for all confirmed cluster radio galaxies. The shaded portion of the histogram represents those values for which the FIR flux, and hence the $q$ value, is an upper limit. The strong peak at $q\sim2.3$ results from the standard FIR-radio correlation, and consequently the majority of such galaxies are normal star-forming galaxies. Galaxies with $q\lesssim2$ are normally AGN including LINERs and Seyferts, while those with $q\lesssim1$ are powerful radio sources with elliptical hosts. \label{fig3}}

\clearpage

\begin{deluxetable}{c c c c c c c c}
\tablecaption{Adopted Cluster Parameters \label{tbl-1}}
\tabletypesize{\small}
\tablewidth{0pt}
\tablehead{
\colhead{Cluster} & \colhead{Richness} & \colhead{RA} & \colhead{Dec} &
\colhead{z} & \colhead{$\sigma$ [km s$^{-1}$]} & \colhead{$m_{zero}$} & 
\colhead{$A_R$}
}
\startdata
A262  & 0 & 01:52:48 & 36:08:00 & 0.0163 & \phn588 & 29.99 & 0.13 \\
A347  & 0 & 02:25:48 & 41:52:00 & 0.0184 & \phn777 & 30.10 & 0.11 \\
A397  & 0 & 02:57:00 & 15:57:00 & 0.0327 & \phn566 & 30.10 & 0.22 \\
A400  & 1 & 02:57:36 & 06:01:00 & 0.0244 & \phn663 & 30.10 & 0.17 \\
A426  & 2 & 03:19:48 & 41:30:42 & 0.0179 & 1324 & 30.10 & 0.30 \\
A539  & 1 & 05:16:36 & 06:27:00 & 0.0284 & \phn629 & 30.10 & 0.28 \\
A569  & 0 & 07:09:18 & 48:37:00 & 0.0201 & \phn327 & 30.10 & 0.18 \\
A634  & 0 & 08:14:36 & 58:02:00 & 0.0265 & \phn391 & 30.10 & 0.05 \\
\enddata

\tablenotetext{a} {Digitized POSS II images not available for this cluster at the time of this writing. POSS I images have been used instead.}

\tablecomments{Column legend --- (1) ACO cluster designation; (2) ACO richness class; (3-4) Adopted cluster center position, in J2000 coordinates; (5) Adopted cluster redshift; (6) Adopted cluster velocity dispersion, from Struble \& Rood; (7) Zero point magnitude assumed for POSS II photometry of the cluster; (8) Average extinction assumed for POSS II photometry.}

\end{deluxetable}

\clearpage

\begin{deluxetable}{c c c c c c c }
\tablecaption{Extended Radio Sources \label{tbl-2}}
\tabletypesize{\footnotesize}
\tablewidth{0pt}
\tablehead{
\colhead{Source} & \colhead{Cluster} & \colhead{Separation} &
\colhead{$S_{1.4GHz}$} & \colhead{FIRST?} & \colhead{Accept?} & 
\colhead{Other Name}
}
\startdata
022259+430042\tablenotemark{a} & A347 & $>$99.9 & $<$8577.\phn & N & Y & UGC01837 \\
022308+412211\tablenotemark{b} & A347 & \phm{$>$}31.5 & \phn\phn$<$20.5 & N & Y & UGC01840 \\
022311+412204\tablenotemark{b} & A347 & \phm{$>$}49.4 & \phn\phn$<$20.5 & N & Y & UGC01840 \\
025654+164844\tablenotemark{a} & A397 & \phm{$>$}50.1 & \phn\phn$<$18.9 & N & Y & --- \\
025742+060128 & A400 & \phm{$>$}71.0 & \phm{$>$}5836.\phn & N & Y & 3C75\tablenotemark{c} \\
125635+281628 & A1656 & \phm{$>$}41.0 & \phm{$>$}\phn201.5 & Y & N & --- \\
125818+290739 & A1656 & \phm{$>$}73.9 & \phm{$>$}\phn\phn\phn8.6 & Y & N\tablenotemark{d} & --- \\
125923+275441 & A1656 & \phm{$>$}25.3 & \phm{$>$}\phn445.6 & Y & Y & NGC4869\tablenotemark{e} \\
\enddata

\tablenotetext{a}{Nearby stronger radio source. Accepted on account of radio morphology in contour image.}

\tablenotetext{b}{Pair of galaxies associated with one radio source. The contour images suggests both are radio sources.}

\tablenotetext{c}{See, e.g., Ledlow \& Owen (1995a)}
\notetoeditor{The citation for note (c) is \verb=\citet{leda1995}=}
\tablenotetext{d}{Elliptical galaxy with weak H$\alpha$ emission; FIRST image suggests radio emission associated with the galaxy is mainly from a background source.}

\tablenotetext{e}{See, e.g., Sijbring (1993)}
\notetoeditor{The citation for note (e) is \verb=\citet{sijb1993}=}
\tablecomments{Column legend --- (1) Source name from J2000 coordinates; (2) ACO cluster in which radio galaxy appears; (3) Separation between optical position and radio position as specified in the NVSS catalog; (4) Radio flux at 1.4GHz, determined directly from the NVSS image. If the radio source lies within a stronger extended source, an upper limit is indicated; (5) Indication of whether a FIRST image of the source is currently available; (6) Indication of whether the source is accepted in the list of cluster radio galaxy candidates, based on the FIRST image. If no FIRST image is available and only the NVSS image is used, the source is accepted. See Figure 1; (7) Common name of the radio galaxy, and reference where applicable.}

\end{deluxetable}

\clearpage

\begin{deluxetable}{c c c c c c c}
\tablecaption{Spectroscopy Observing Runs \label{tbl-3}}
\tabletypesize{\footnotesize}
\tablewidth{0pt}
\tablehead{
\colhead{Dates} & \colhead{Observatory} & \colhead{Grating} & \colhead{Blaze} &
\colhead{Dispersion} & \colhead{Slit Width} & \colhead{Wavelength Range} 
}
\startdata
12/8/98, 12/24/98 & APO & Med-B & 4861 & 3.18 & 2$\arcsec$ & 3850 - 5275 \\
                  &     & Med-R & 6563 & 3.55 & 2$\arcsec$ & 5250 - 7250 \\
1/9/99 - 1/11/99  & APO & Med-B & 4861 & 3.18 & 2$\arcsec$ & 3825 - 5250 \\
                  &     & Med-R & 6563 & 3.53 & 2$\arcsec$ & 5275 - 7275 \\
4/9/99 - 4/11/99  & APO & Med-B & 4861 & 3.18 & 2$\arcsec$ & 3750 - 5250 \\
                  &     & Med-R & 6563 & 3.54 & 2$\arcsec$ & 5300 - 7300 \\
5/8/99 - 5/12/99  & KPNO & 09 & 4000 & 2.40 & 2$\arcsec$ & 3700 - 8000 \\
12/4/99 - 12/10/99 & KPNO & 09 & 4000 & 2.43 & 2$\arcsec$ & 3600 - 7900 \\
\enddata

\tablecomments{Column legend --- (1) Calendar dates of scheduled observations. Note that APO time is scheduled in half-night blocks so each date corresponds to half a night of scheduled observing; (2) Observatory where observations were performed. APO signifies observations at the Apache Point Observatory with the 3.5-meter telescope and the Double Imaging Spectrograph, and KPNO signifies observations made at the Kitt Peak National Observatory on the 2.1-meter telescope with the GoldCam spectrograph; (3) Choice of gratings. Note that for APO, the incident light is split into a blue and a red spectrum by a dichroic. The gratings and detector characteristics differ for the two systems; (4) Blaze angle of grating, in Angstroms. APO's gratings are believed to have blaze angles corresponding to the Balmer lines; (5) Resolution of resulting spectra, in $\mbox{\AA} ~pixel^{-1}$; (6) Slit width, in arcseconds; (7) Wavelength range of resulting spectra, in Angstroms. Note that for the APO red spectra, the range extends beyond that quoted but we have truncated it in the table as our sensitivity function is fit over this this range. In the case of the KPNO spectra, the chip size allows a greater wavelength range but instrument focus degrades at the extremes.}

\end{deluxetable}

\clearpage

\begin{deluxetable}{c c c c c}
\tablecaption{Measured Radio Galaxy Velocities \label{tbl-4}}
\tabletypesize{\small}
\tablewidth{0pt}
\tablehead{
\colhead{Source} & \colhead{Observing Run} & \colhead{z} & \colhead{Error} &
\colhead{Emission Lines} 
}
\startdata
014736+373723 & KPNO-2 & 0.0744 & 0.0002 & (H$\alpha$, [NII]) \\
015040+372308 & KPNO-2 & 0.0737 & 0.0002 & --- \\
015129+360356 & KPNO-2 & 0.0178 & 0.0001 & H$\alpha$, H$\beta$, [NII], [SII] \\
015240+361016 & KPNO-2 & 0.0186 & 0.0002 & (H$\alpha$, [NII]) \\
015246+360907 & KPNO-2 & 0.0171 & 0.0004 & (H$\alpha$, [NII]) \\
015254+360311 & KPNO-2 & 0.0206 & 0.0001 & H$\alpha$, H$\beta$, [NII] \\
015417+341855 & KPNO-2 & 0.0859 & 0.0002 & --- \\
015628+364807 & KPNO-2 & 0.0198 & 0.0002 & H$\alpha$, H$\beta$, [NII], [SII] \\
\enddata

\tablenotetext{a}{Due to mis-alignment of the slit, this is an off-nuclear spectrum. Nuclear emission lines may be present.}
\tablenotetext{b}{Outside radial extent of sample; included as serendipitous finding of an extended radio galaxy.}
\tablenotetext{c}{This appears to be an interacting system, with the radio emission centered on emission line object.}

\tablecomments{Column legend --- (1) Source name, from J2000 coordinates; (2) Observing run during which spectrum was collected. `APO' represents data collected at Apache Point Observatory, with the attached number indicating the observing run (1 = December 1998, 2 = January 1999, 3 = April 1999). `KPNO' represents data collected at Kitt Peak National Observatory, with the attached number indicating the observing run (1 = May 1999, 2 = December 1999); (3) Heliocentric redshift of galaxy; (4) Error in redshift measurement; (5) Where emission lines are present, the galaxy's redshift has been determined from them and the identified lines are indicated here. If no emission lines are present (`---'), the velocity has been determined via fourier cross correlation of the galaxy's absorption lines. In the event that emission lines are present but weak, the velocity has been determined via fourier cross correlation and identified lines are listed in parentheses.}

\end{deluxetable}

\clearpage

\begin{deluxetable}{c c c c c c c c c c}
\tablecaption{Cluster Radio Galaxy Data \label{tbl-5}}
\tabletypesize{\small}
\tablewidth{0pt}
\tablehead{
\colhead{Cluster} & \colhead{RA} & \colhead{Dec} &
\colhead{z} & \colhead{m} & \colhead{$S_{1.4GHz}$} & 
\colhead{$S_{60\mu m}$} & \colhead{$S_{100\mu m}$} &
\colhead{q} & \colhead{Common} \\
\colhead{} & \colhead{(J2000)} & \colhead{(J2000)} &
\colhead{} & \colhead{} & \colhead{[mJy]} & 
\colhead{[Jy]} & \colhead{[Jy]} &
\colhead{} & \colhead{Name}
}
\startdata
A262 & 01:41:15.6 & 34:48:40.3 & 0.0170\tablenotemark{a} & 13.0 & 5.8 & 0.788 & 1.443 & 2.30 & --- \\*
     & 01:44:35.5 & 37:41:46.1 & 0.0189\tablenotemark{a} & 12.7 & 18.9 & 1.878 & 4.595 & 2.22 & NGC0662 \\*
     & 01:44:38.3 & 34:39:35.5 & 0.0189\tablenotemark{a} & 13.9 & 3.7 & 0.362 & 0.807 & 2.20 & --- \\*
     & 01:46:22.7 & 36:27:35.7 & 0.0150\tablenotemark{a} & 12.3 & 3.4 & 0.750 & 2.048 & 2.60 & NGC0668 \\*
     & 01:46:57.0 & 34:46:19.3 & 0.0140\tablenotemark{a} & 13.2 & 4.2 & 0.496 & 1.191 & 2.30 & VV701 \\*
     & 01:50:41.1 & 33:44:25.8 & 0.0187\tablenotemark{a} & 14.1 & 4.0 & 0.548 & 1.017 & 2.31 & MRK1008 \\*
     & 01:50:44.3 & 35:17:03.2 & 0.0138\tablenotemark{a} & 12.1 & 10.3 & 1.286 & 2.783 & 2.30 & NGC0688 \\*
     & 01:51:29.3 & 36:03:56.5 & 0.0178     & 12.9 & 19.4 & 0.923 & 2.736 & 1.95 & UGC01319 \\
\enddata
\end{deluxetable}

\clearpage

\begin{deluxetable}{c c c c c c c c c c}
\tablecaption{Non-Cluster Radio Galaxy Data \label{tbl-6}}
\tabletypesize{\small}
\tablewidth{0pt}
\tablehead{
\colhead{Cluster} & \colhead{RA} & \colhead{Dec} &
\colhead{z} & \colhead{m} & \colhead{$S_{1.4GHz}$} & 
\colhead{$S_{60\mu m}$} & \colhead{$S_{100\mu m}$} &
\colhead{q} & \colhead{Common} \\
\colhead{} & \colhead{(J2000)} & \colhead{(J2000)} &
\colhead{} & \colhead{} & \colhead{[mJy]} & 
\colhead{[Jy]} & \colhead{[Jy]} &
\colhead{} & \colhead{Name}
}
\startdata
A262 & 01:58:04.5 & 34:17:58.8 & 0.0476\tablenotemark{a} & 13.4 & 34.6 & 0.105\tablenotemark{e} & 0.730\tablenotemark{d} & 0.99 & --- \\
     & 01:47:35.8 & 37:37:23.3 & 0.0744     & 14.0 & 8.5 & 0.160\tablenotemark{d} & 0.570\tablenotemark{d} & 1.59 & --- \\
     & 01:50:40.4 & 37:23:07.6 & 0.0737     & 13.8 & 128.0 & 0.126\tablenotemark{e} & 0.234\tablenotemark{e} & 0.17 & --- \\
     & 01:54:17.1 & 34:18:54.6 & 0.0859     & 13.6 & 23.3 & 0.210\tablenotemark{d} & 0.530\tablenotemark{d} & 1.19 & --- \\
     & 01:59:58.3 & 37:23:31.2 & 0.0489     & 14.0 & 7.6 & 0.084\tablenotemark{e} & 0.225\tablenotemark{e} & 1.29 & --- \\
A347 & 02:13:58.8 & 41:20:29.5 & 0.0874     & 14.2 & 2.9 & 0.170\tablenotemark{d} & 0.910\tablenotemark{d} & 2.19 & --- \\
     & 02:16:12.0 & 42:49:19.1 & 0.0457\tablenotemark{a} & 13.1 & 3.2 & 0.480\tablenotemark{d} & 1.300\tablenotemark{d} & 2.43 & UGC01743 \\
     & 02:24:29.5 & 40:52:12.5 & 0.0434     & 13.0 & 3.3 & 0.730\tablenotemark{d} & 1.640\tablenotemark{d} & 2.55 & UGC01855 \\
\enddata

\tablenotetext{a}{Redshift found in NED.}
\tablenotetext{b}{Resolved radio source. The flux was determined directly from the NVSS image.}
\tablenotetext{c}{Extended radio source (see Table 2).}
\tablenotetext{d}{IRAS detection, with data from {\it xscanpi} (see text).}
\tablenotetext{e}{Non-detection. The value quoted is a 3$\sigma$ upper limit based on the local rms reported by {\it xscanpi}.}
\tablenotetext{f}{Overlay shows radio emission originates from background galaxy and not the cluster member.}
\tablenotetext{g}{Magnitude contaminated by star.}

\tablecomments{Column legend --- (1) ACO cluster designation; (2) Right Ascension of radio galaxy, based on the optical position in J2000 coordinates; (3) Declination of radio galaxy, based on the optical position in J2000 coordinates; (4) Heliocentric redshift of the galaxy, where a flag indicates the redshift was found in NED; (5) Apparent Gunn-Oke magnitude of the galaxy, derived from the POSS II (see text); (6) 1.4GHz radio flux of the galaxy, taken from the NVSS. For sources contained within larger extended radio emission (see Table 2), the flux is an upper limit and denoted with a less than sign; (7) \& (8) IRAS 60 $\mu$m flux and 100 $\mu$m flux, if available. Unless noted otherwise, the flux is taken from either the IRAS Faint Source Catalog or Point Source Catalog; (9) $q$ value for the galaxy, derived from columns (6) -- (8). Note that the reported $q$ value may represent an upper limit (see flags for columns (7) \& (8)); (10) Name of galaxy in other common catalogs, with preference given to the NGC and UGC catalogs.}

\end{deluxetable}

\clearpage

\begin{deluxetable}{c c c c c c}
\tablecaption{Absolute Quantities \label{tbl-7}}
\tablewidth{0pt}
\tablehead{
\colhead{Cluster} & \colhead{RA} & \colhead{Dec} &
\colhead{$M_R$} & \colhead{$P_{1.4GHz}$} & \colhead{$L_{FIR}$} \\
\colhead{} & \colhead{(J2000)} & \colhead{(J2000)} &
\colhead{} & \colhead{$\log$[W Hz$^{-1}$]} & \colhead{log[L$_{\sun}]$}
}
\startdata
A262 & 01:41:15.6 & 34:48:40.3 & -21.2 & 21.5 & 9.8 \\
     & 01:44:35.5 & 37:41:46.1 & -21.5 & 22.0 & 10.2 \\
     & 01:44:38.3 & 34:39:35.5 & -20.3 & 21.3 & 9.5 \\
     & 01:46:22.7 & 36:27:35.7 & -21.9 & 21.2 & 9.8 \\
     & 01:46:57.0 & 34:46:19.3 & -21.0 & 21.3 & 9.6 \\
     & 01:50:41.1 & 33:44:25.8 & -20.1 & 21.3 & 9.6 \\
     & 01:50:44.3 & 35:17:03.2 & -22.1 & 21.7 & 10.0 \\
     & 01:51:29.3 & 36:03:56.5 & -21.3 & 22.0 & 9.9 \\
\enddata

\tablenotetext{a}{Object IR flux densities derived using {\it xscanpi}. The source was detected at 60$\mu$m or 100$\mu$m, but not both. The FIR luminosity is estimated using a flux density in the non-detected band equal to three times the rms noise.}

\tablenotetext{b}{Object not detected at either 60$\mu$m or 100$\mu$m. Upper limit based on three times the rms noise at 60$\mu$m or 100$\mu$m, as determined from {\it xscanpi}.}

\tablenotetext{c}{Photometry corrupted by presence of a star inside the aperture.}

\tablenotetext{d}{Pair of galaxies. The NVSS flux likely represents the combination of the two galaxies, and the corresponding power is consequently an upper limit. The galaxy pair may be seen in Figure 1.}

\tablecomments{Column legend --- (1) ACO cluster designation; (2) RA of cluster radio galaxy; (3) Dec of cluster radio galaxy; (4) Absolute magnitude of cluster radio galaxy, computed for Gunn-Oke aperture; (5) Logarithm of radio power at 1.4 GHz; (6) Logarithm of FIR luminosity, in units of solar luminosities.}

\end{deluxetable}

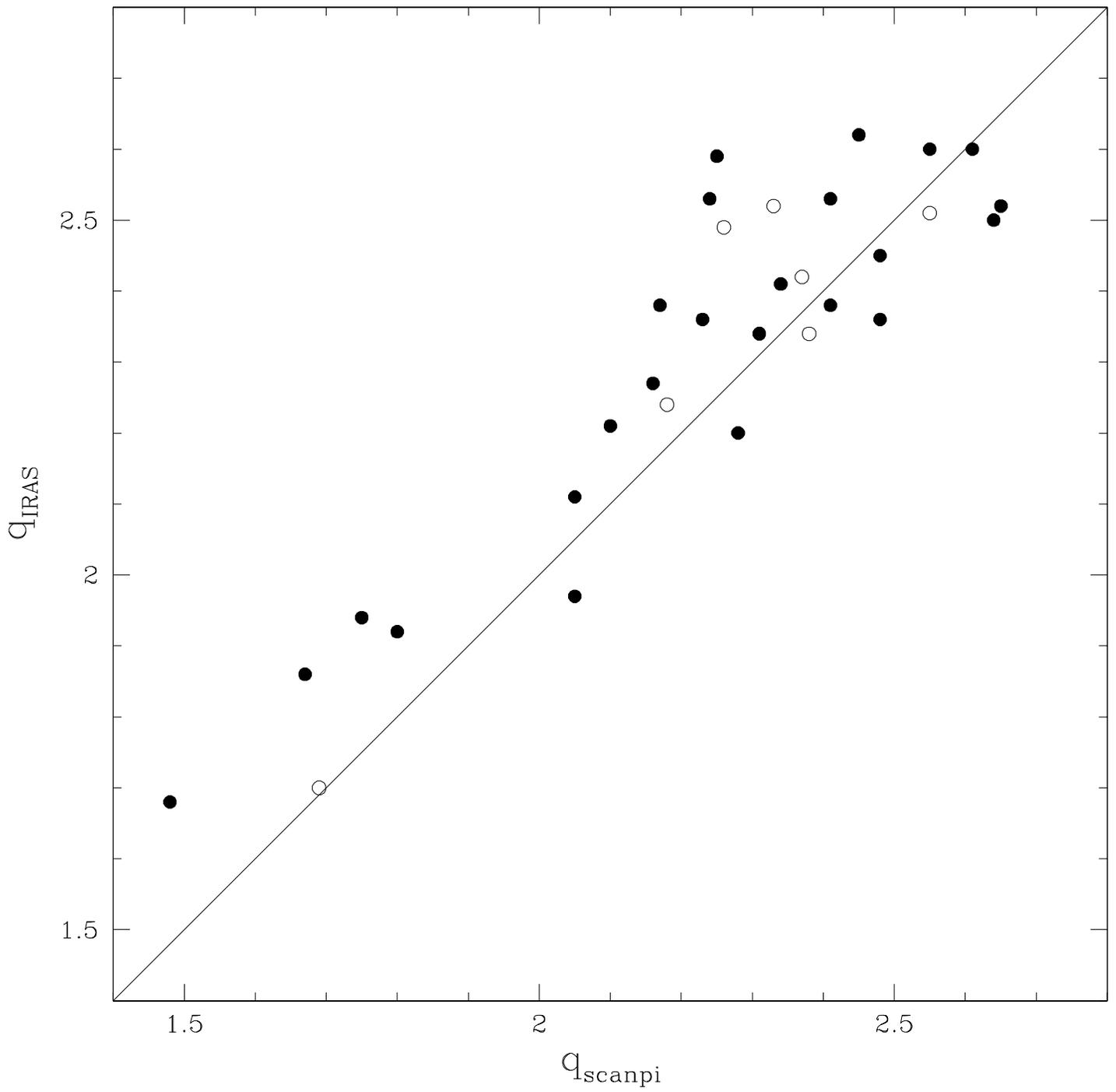
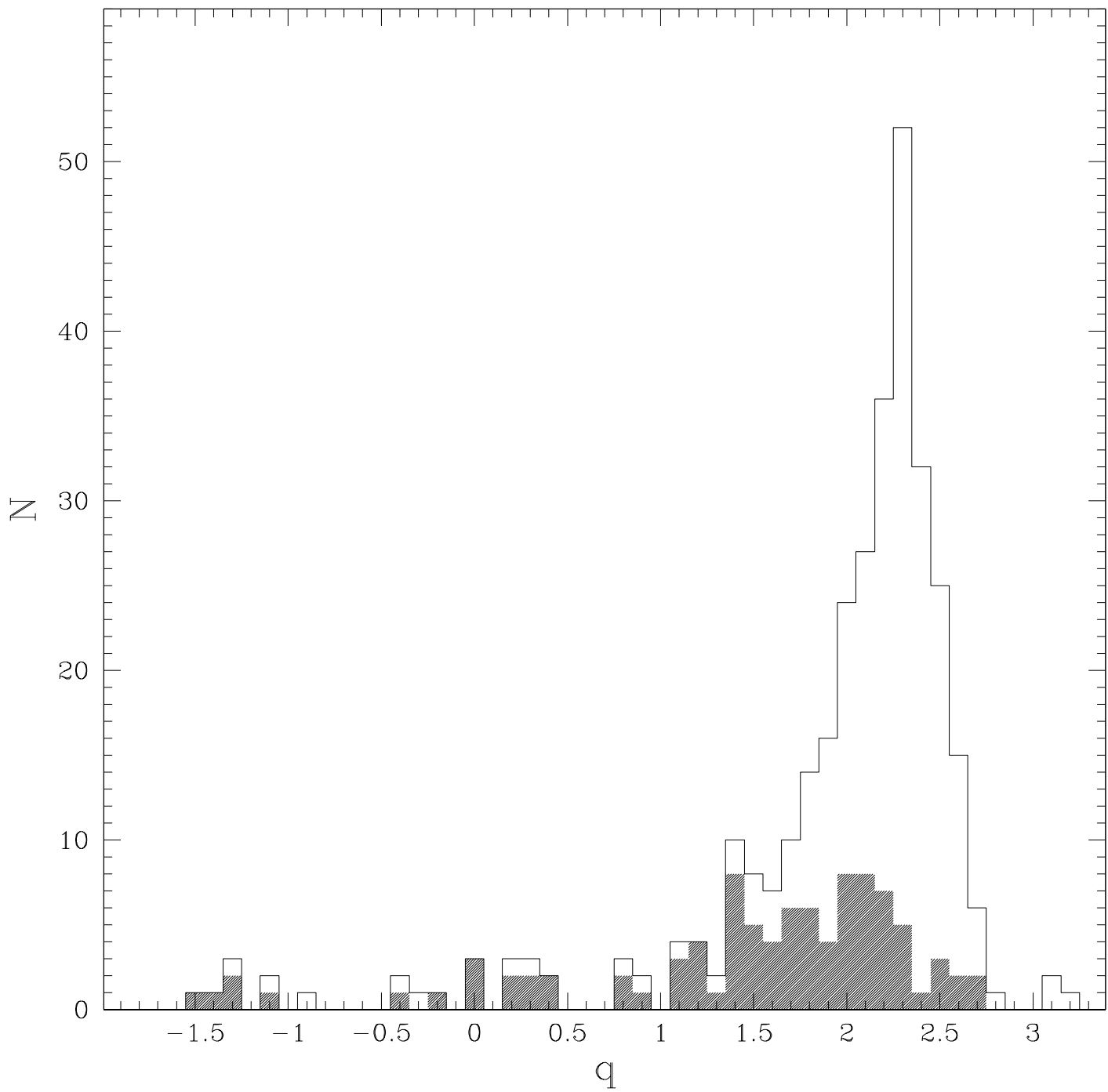
\end{document}